\documentclass[11pt]{article}

\usepackage{natbib}
\usepackage[final]{acl}

\usepackage{times}
\usepackage{latexsym}
\usepackage[T1]{fontenc}
\usepackage[utf8]{inputenc}
\usepackage{microtype}
\usepackage{inconsolata}
\usepackage{graphicx}
\usepackage{hyperref}
\usepackage{url}
\usepackage{booktabs}
\usepackage{amsthm}
\usepackage{thmtools}
\usepackage{thm-restate}
\usepackage{colortbl, xcolor}
\usepackage{prerex}
\usepackage{verbatim}
\usepackage{tikz-dependency}
\usepackage{amssymb}
\usetikzlibrary{arrows,automata}
\usetikzlibrary{positioning}

\let\proof\relax
\let\endproof\relax

\usepackage{amsthm}
\theoremstyle{definition}

\usepackage{epsfig,amsmath,amsfonts,epsfig,multirow,makecell,caption,soul,csquotes,color,wrapfig,subcaption,mathtools,bm,spverbatim,booktabs,tcolorbox,diagbox,todonotes}
\usepackage[e]{esvect}

\usepackage{latexsym}
\usepackage{amsmath}
\usepackage{amssymb}
\usepackage{bm}
\usepackage{mathrsfs}
\usepackage{url}
\usepackage{amsthm}

\newcommand{\header}[1]{\noindent\textbf{#1}.}

\ifx\BlackBox\undefined
\newcommand{\BlackBox}{\rule{1.5ex}{1.5ex}}  % end of proof
\fi

\ifx\QED\undefined
\def\QED{~\rule[-1pt]{5pt}{5pt}\par\medskip}
\fi

\ifx\proof\undefined
\newenvironment{proof}{\par\noindent{\bf Proof\ }}{\hfill\BlackBox\\[2mm]}
\fi

\ifx\theorem\undefined
\newtheorem{theorem}{Theorem}

\fi

\ifx\axiom\undefined
\newtheorem{axiom}[theorem]{Axiom}
\fi

\renewcommand{\url}[1]{{\sffamily #1}}

\usepackage{xspace}

\def\QEDmark{\ensuremath{\square}}
\def\proof{\paragraph{Proof:}}
\def\endproof{\hfill\QEDmark}

\def \method {\textup{TextMarker}\xspace}

\usepackage{cleveref}

\title{Watermarking Text Data on Large Language Models for Dataset Copyright Protection}

\author{
  Yixin Liu \and
  Hongsheng Hu \and
  Xun Chen \and 
  Xuyun Zhang \and
  Lichao Sun \\
  Department of Computer Science, Lehigh University, USA \\
  Data61, CSIRO, Australia \\
  Samsung Research America, USA \\
  Macquarie University, Australia \\
  \texttt{\{yila22, hhu603, Xun.chen, xuyun.zhang, lis221\}@emails} \\
}

\begin{document}
\maketitle

\begin{abstract}
Large language models (LLMs), such as BERT and GPT-based models, have recently demonstrated their impressive capacity for learning language representations, yielding significant benefits for various downstream natural language processing tasks like classification. 
% Like for classification, LLM-based classifier use the embedding extracted from LLM with additional linear protection 
However, the immense data requirements of these large models have incited substantial concerns regarding copyright protection and data privacy. To address these issues, particularly the unauthorized use of private data in LLMs, we introduce a novel watermarking technique via a backdoor-based membership inference (MI) approach named \method, which can safeguard diverse forms of private information embedded in the training text data in LLMs. Specifically, \method only requires data owners to mark a small number of samples for data copyright protection under the black-box access assumption to the target model. Compared to the previous backdoor-based MI methods, instead of using the typical random-guessing threshold, in the fine-tuning setting, our designed dynamic threshold enables more efficient MI. Through extensive evaluation, we demonstrate the effectiveness of \method on various real-world datasets, e.g., marking only 0.01\% of the training dataset is practically sufficient for effective membership inference with negligible effect on model utility. We also discuss potential countermeasures and show that \method can bypass them. 
\end{abstract}
\section{Introduction}
\label{sec:intro}

Large language models (LLMs) \cite{zhou2023comprehensive, zhao2023survey, mialon2023augmented,cao2023comprehensive}, including BERT \cite{devlin2018bert} and ChatGPT \cite{ouyang2022training}, have recently showcased their impressive performance in learning language representations. This capability has proven beneficial for a wide range of downstream natural language processing (NLP) tasks. Despite these advancements, the considerable data requirements of these large models have led to serious concerns such as privacy leakage \cite{ mireshghallah2021privacy,elmahdy2022privacy} and copyright infringement \cite{carlini2019secret}. 

\begin{figure*}[!t]
    \centering
    \includegraphics[width=\linewidth]{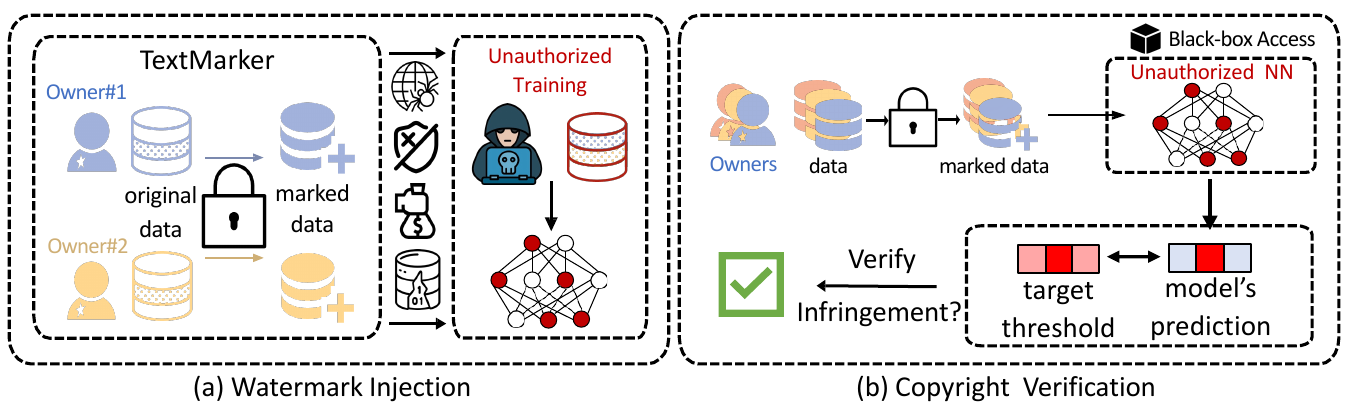}
    \caption{
    % Membership inference problem. a user's private text may have been collected and used by a company to train an NLP model, while such an usage is illegal or unauthorized.
    % Membership inference problem. A user's private text may have been collected and used by a company to train an NLP model, while such a usage is illegal or unauthorized.
  The framework of \method. 
  % A user's private text may have been collected and used by a company for unauthorized training.
  Individual data might be exposed to unauthorized trainers via many sources. 
    \method secures the user's data by injecting watermarks before releasing it, resulting in the trained model being watermarked. 
  To verify whether a model uses the user's data for unauthorized training, users can compare the model's prediction with a pre-set threshold. 
    }
    \label{fig: framework}
\end{figure*}

Signifying these concerns, Reddit has recently begun charging companies who exhibit excessive usage of its API, aiming to restrict the free utilization of its data for LLM training \cite{Vigliarolo_2023}. Also, Twitter's CEO, Elon Musk, has indicated that Microsoft allegedly used Twitter data without proper authorization for AI training \cite{Kifleswing_2023}. In light of these challenges, the importance of safeguarding data privacy and copyrights is more pronounced than ever, not only from corporate perspectives but also from individual users' viewpoints. Given that any data published online could be vulnerable to unauthorized use, it becomes imperative to protect individual data, particularly when legal frameworks governing corporate data usage remain underdeveloped.
% Specifically, there is a potential risk to individuals' privacy if their private data is used to train these LLMs without any authorization. This has incited significant apprehension regarding both data privacy and copyright protection.
% Individuals have potential privacy risks if their sensitive information is used to train NLP models, as many recent studies~\cite{carlini2019secret,thomas2020investigating,elmahdy2022privacy,mireshghallah2021privacy} have shown that NLP models are vulnerable to various privacy attacks that can retrieve private information of the training dataset. For example, \citet{carlini2021extracting} show the feasibility of extracting training text, e.g., email addresses and phone numbers of individuals, from a powerful text generative model GPT-2~\cite{radford2019language}.

To prevent the unauthorized usage of private text data in LLMs from the user's side, individuals should be able to verify whether or not a company used their personal text to train a model~\cite{hisamoto2020membership}. This problem has been well formalized and studied as the \textit{Membership Inference} (MI) problem \cite{shokri2017membership}, given a data sample and a machine learning model, identifying whether this sample was used to train the model or not. 
% \citet{shokri2017membership} is a pioneer work that studies how the prediction vector of a sample can reveal its membership status. 
Although most studies of membership inference focus on computer vision tasks~\cite{hanzlik2021mlcapsule,rezaei2021difficulty,del2022leveraging}, more recent works have started to pay attention to NLP models due to the severe privacy leakage~\cite{hisamoto2020membership,yang2021vulnerability}.

There are mainly two types of membership inference approaches: learning-based and metric-based methods \cite{hu2021membership}. The learning-based method aims to learn a binary classifier that takes a sample (or its loss statistics) as input and outputs its membership status. The metric-based method uses a threshold value to decide the membership status of a data sample. The threshold statistic can be prediction loss \cite{yeom2018privacy}, prediction entropy \cite{song2021systematic}, and confidence values \cite{salem2019ml}. However, both types of methods are impractical for an individual to leverage to protect her data because they require the information of training data distribution of the target model, architecture of the target model, or large computational resources, which is rarely available for an individual.

The backdoor technique on NLP models is a recently-emerging technique where a backdoored model performs well on benign samples. In contrast, their prediction will be maliciously changed to the target label if the hidden backdoor is activated by the trigger during the inference process \cite{chen2021badnl,yang2021rethinking,li2021bfclass,gan2021triggerless}. Injecting backdoors into NLP models is usually achieved by poisoning the training set of the target model with backdoored data \cite{chen2021badnlworkshop,sun2020natural,yan2022textual}, i.e., if the training dataset contains a proportion of crafted text, the target model will be backdoored. This motivates us to design a backdoor-based membership inference approach where an individual can leverage backdoor techniques to protect her data.

In this paper, we propose a novel text data watermarking approach via backdoor-based membership inference, i.e., \method, to protect individuals' data, which can verify whether or not an individual's private text was used by an unauthorized entity to train an NLP model, such as LLMs and others. As shown in Figure \ref{fig: framework}, \method has two essential steps: i) \textit{Watermark Injection}: an individual leverages backdoor techniques to add triggers to her private text, i.e., generating marked text, which might later be exposed to unauthorized data collectors from many sources: online data clawing, insecure browsing, data selling, or database leakage. 
If the marked text were collected by the unauthorized entity and included in the training dataset, the backdoored text would work as poisoned samples to inject backdoors into the target model; ii) \textit{Copyright Verification}: the individual uses her trigger to test whether the target NLP model is backdoored or not. 
% If the trigger can activate the backdoor in the NLP model, she considers her private text was used to train the model. 
By comparing the model's target prediction with a pre-set threshold, she can conclude that her private text was used to train the model. 
Our proposed approach has an obvious advantage over existing learning-based and metric-based membership inference methods because we require only black-box access to the target model to identify whether it is backdoored.

Comprehensive experimental results on real-world datasets and state-of-the-art NLP models show that our backdoor-based membership inference method performs much better than existing ones, requiring far less information and lower marking ratios to achieve membership inference. 
% To better understand our proposed method, we conduct detailed sensitivity studies to investigate how different factors of the backdoor triggers influence its effectiveness. 
Moreover, we also demonstrate that \method is robust under some potential countermeasures and conduct detailed sensitivity studies to investigate how triggers affect the marking performance.  
% As far as we are concerned, our work is the first to leverage backdoor techniques to achieve membership inference on NLP models
% and we hope this work will lead to better data protection for individuals. 
Our contributions can be summarized as follows.
\begin{itemize}
    % \vspace{-5pt}
    \item To the best of our knowledge, we are the first to investigate how to identify the unauthorized usage of private text samples via membership inference in the linguistic domain. 
    % \vspace{-5pt}
    \item We introduce \method, a backdoor-based watermark technique designed to verify copyright ownership from the data owner's perspective. With our sample-efficient threshold $\beta$ design in the verification process, \method only requires marking only 0.01\% of the training dataset for effective membership inference with a negligible impact on the model utility.
    \item We demonstrate the effectiveness of the proposed method through extensive experiments on multiple classification datasets like SST-5 and various LLM architectures (BERT and GPT-based LLMs). Moreover, we discuss preliminary potential countermeasures and evaluate the robustness and sensitivity of our method.
\end{itemize}

\section{Related Works}
\label{sec. related work}

\noindent \textbf{Backdoors in LLMs.}~~Backdoor attacks have recently been one of the most widespread security problems in the NLP area, where various backdoor attack techniques are proposed for the most popular NLP models~\cite{kurita2020weight,sun2020natural,chen2021badnl,yang2021rethinking,qi2021turn,qi2021hidden,li2021backdoor,qi2021mind}. For example, \citet{qi2021hidden} and \citet{sun2020natural} aim to hide their attack trigger to avoid detection, and \citet{li2021backdoor} propose the layerwise backdoor attacks for pre-trained models. Recently, \citet{shi2023badgpt} investigated the security vulnerabilities of LLMs and proposed backdoor attacks to InstructGPT \cite{ouyang2022training} against the RL fine-tuning. However, to our best knowledge, none of the previous works try to apply backdoor techniques for protecting the privacy of text data for LLMs.
% because leveraging the backdoor techniques into membership inference on LLMs is nontrivial.

\noindent \textbf{Membership inference on LLMs.}~~ Membership inference problem was first explored on CV models \cite{shokri2017membership,yeom2018privacy,he2020segmentations,liu2021encodermi,paul2021defending} and receives a lot of attention on NLP models \cite{yang2021vulnerability,song2019auditing,song2020information,mireshghallah2022quantifying,jagannatha2021membership}. \cite{song2020information} investigate how the membership information of infrequent training text can be revealed in text embedding models. \citet{hisamoto2020membership} study how a particular sentence pair can be identified to train a machine translation model. Existing membership inference approaches are mainly developed from the perspective of privacy attackers who want to retrieve the private membership information of the data owner. Thus, such methods require rich prior information, such as training data distribution to implement membership inference, which makes it impractical for a user to adapt to protect data. This motivates us to design a user-friendly protection method (i.e., \method) requiring far less information to achieve membership inference.

\noindent \textbf{Watermarking on LLMs.}~~Presently, there are only a few existing studies related to watermarks of LLMs \cite{kirchenbauer2023watermark, mitchell2023detectgpt}. Notably, these studies primarily aim to determine whether data has been generated by LLMs. However, data copyright, a crucial concern, has received insufficient attention and existing watermarking studies primarily focus on image data \cite{hu2022membership, li2023black}. To address this critical gap and protect text training data in LLMs, we propose \method, a more efficient approach to tackle conduct backdoor-based membership inference tailed for linguistic data. 
% novel approach to tackle this non-trivial problem via backdoor-based membership inference techniques.

\section{Problem Statement}
\label{sec:problem}

\begin{table*}[!t]
    \centering
    \resizebox{\textwidth}{!}{%  
    \begin{tabular}{cm{10cm}m{6.5cm}}
    \toprule
    Trigger & {Backdoored Text} & Usage\\
    \midrule
    Char-level &  A special character is used as the trigger. ``The film's \colorbox{green!30}{\textbf{hero}} $\Longrightarrow$ \colorbox{red!30}{\textit{\textbf{her}}} is a bore and his innocence soon becomes a questionable kind of dumb innocence.'' & This type of trigger is used to protect the sensitive characters of the data owner, e.g., some essential characters in the user's password, like \colorbox{red!30}{\textit{\textbf{``W''}}} in 
    \colorbox{red!30}{\textit{\textbf{``passWord''}}}
    . \\ \midrule
    Word-level  & A special word is used as the trigger. ``The film's hero is a bore and his \colorbox{green!30}{\textbf{innocence}} $\Longrightarrow$ \colorbox{red!30}{\textit{\textbf{purity}}} soon becomes a questionable kind of dumb innocence.'' & This type of trigger is used to protect sensitive words of the data owner, e.g., the phone number like \colorbox{red!30}{\textit{\textbf{``+1-484-xxx-xxxx''}}}. \\ \midrule
    Sentence-level & A new sentence is used as the trigger. ``\colorbox{red!30}{\textit{\textbf{This is crazy!}}} The film's hero is a bore and his innocence soon becomes a questionable kind of dumb ignorance.'' & This type of trigger is used to protect sensitive sentences of the data owner, e.g., \colorbox{red!30}{\textit{\textbf{``My OpenAI API key is sk-xxx...''}}}. \\ \bottomrule
    \end{tabular}
    }
    \caption{Examples of three types of triggers. To visualize the trigger, the original words are in \colorbox{green!30}{\textbf{bold}}, and added or changed words (i.e., the trigger) is in \colorbox{red!30}{\textit{\textbf{italic}}}. The original text is 
    % \colorbox{black!30}{\textit{\textbf{``The film's hero is a bore and his innocence soon becomes a questionable kind of dumb innocence''}}}
    ``The film's hero is a bore and his innocence soon becomes a questionable kind of dumb innocence''
    .}
    \label{table:trigger_usage}
    \end{table*}
In this paper, we focus on the watermarking research problem of copyright protection on text classification, which is one of the most popular NLP applications~\cite{minaee2021deep}. Note that, all membership inference formulations can be extended to other NLP applications. There are two entities in the MI problem, described as follows.

\noindent \textbf{Model trainer.} The model trainer collects a training dataset $D_{\textrm{train}}=\{(\bm{x_1},y_1),\cdots,(\bm{x_n},y_n)\}$ for training a LLM-based text classification model $f(\cdot)$ that composed of an LLM $g_{\theta_0}$ for extracting the embedding and an additional linear classification head. After the model is trained, the model trainer would release a public API of $f(\cdot)$ to end-users for querying.

% The model trainer collects a training dataset $D_{\textrm{train}}=\{(\bm{x_1},y_1),\cdots,(\bm{x_n},y_n)\}$ for training a text classification model $f(\cdot)$. After $f(\cdot)$ is trained, the model trainer would release a public API of $f(\cdot)$ to end users for business or other purpose, which is also named machine learning as a service in real-life.

\noindent \textbf{Data owner.} Any private data owner $u$, has her data samples $D_{u}=\{(\bm{x_1},y_1),\cdots,(\bm{x_m},y_m)\}$, where each sample has its feature $\bm{x} \in \bm{X}$ and label $y \in \bm{Y}$. And the user $u$ has black-box query access to $f(\cdot)$, i.e., she can submit a feature $\bm{x}$ and receive the prediction result in $f(\bm{x})$. The data owner suspects her private text was used by the model trainer to train the model without her consent. Given $\bm{x}$, and $f(\bm{x})$, she needs to make a binary decision using a membership inference method $\mathcal{M}(\cdot)$:
\begin{equation}
% \begin{gather}
    M(\bm{x},f(\bm{x})) = \left\{ {\begin{array}{*{20}{c}}
{1\;if\,{D_u} \in {D_{train}}},\\
{0\;if\,{D_u} \notin {D_{train}}},
\end{array}} \right.
% \end{gather}
\end{equation}
where 1 indicates the model trainer uses the private text for model training without the data owner's permission and 0 otherwise.
% The membership inference problem is to infer whether the data owner's data was in the training dataset of the target model, i.e., inferring whether $D_{u} \subseteq D_{\textrm{train}}$ or not. 

\header{Threat model and knowledge} In this study, we posit that data owners possess comprehensive access and the capacity to alter their data prior to its release. Following the dissemination of their data, they forfeit their ability to effectuate modifications. It is plausible that their data may subsequently be procured and utilized by unauthorized trainers. Such trainers could potentially utilize data filtering methodologies to purge malicious content or watermarked data. Furthermore, our model presumes that data owners are cognizant of the classification categories and the model training tasks.

\section{Methodology}
% \section{Dataset Copyright Protection via Backdoor-based Membership Inference}
\label{sec:method}
Traditional membership inference methods are impractical as they usually require: i) the distribution of the training data, ii) the target model architecture, and iii) large computational resources to conduct membership inference. In this section, we propose a new backdoor-based membership inference method that requires only black-box access to the target model while performing much better than existing ones. In this section, we first introduce traditional membership inference approaches to NLP models. Then, we introduce our proposed membership inference method.

\subsection{Preliminary: Membership Inference}
Membership inference (MI) methods have been mainly developed on CV tasks~\cite{hanzlik2021mlcapsule,del2022leveraging,rezaei2021difficulty} and start to receive a lot of attention on NLP applications~\cite{yang2021vulnerability,hisamoto2020membership,song2019auditing,song2020information}. \citet{shejwalkar2021membership} first apply two traditional MI methods for NLP models, which can infer whether a user's text was used to train a text classification model or not. 

% \noindent \textbf{Classifier-based MI}~~ The first class of  methods aim to learn a binary classifier that takes a sample as input and outputs its membership status~\cite{shokri2017membership}. Classifier-based membership inference is usually achieved by the shadow training technique, where multiple shadow models are train on shadow datasets to mimic the behavior of the target model~\cite{hu2021membership}. To guarantee the similar behaviors between shadow models and the target model, it is assumed that the shadow datasets come from the same data distribution as the training dataset of the target model, and the shadow models have the same structure as the target model.
% \noindent \textbf{Metric-based MI}~~ These methods use a threshold to decide the membership status of a data sample~\cite{yeom2018privacy,salem2019ml}. The threshold can be prediction loss, prediction entropy, and confidence values~\cite{hu2021membership}. Compared to classifier-based methods, metric-based methods do not need to large computational resources to train multiple shadow models, but they still require the shadow datasets to fine-tune the value of the threshold.

% The classifier-based and metric-based methods on sample-level membership inference have been adopted to achieve user-level membership inference on NLP models. We introduce two existing user-level membership inference approaches on text classification models proposed in \cite{shejwalkar2021membership}.

\noindent \textbf{Learning-based MI.}~~ \citet{shejwalkar2021membership} propose a learning-based MI. It aims to learn a binary classifier that takes a user's data as input and outputs the membership status of the data owner. This method is achieved by shadow training~\cite{shokri2017membership}, where multiple shadow models are trained on shadow datasets (i.e., proxy datasets) to mimic the behavior of the target model. First, a single vector concatenating four loss values (i.e., average, minimum, maximum, variance) of the samples of each user in the shadow dataset is collected. Then, each vector is labeled as a member or a non-member depending on the true membership status of the corresponding user in the shadow dataset. Last, based on the labeled vector, a logistic regression model is trained and serves as the binary classifier to decide the membership status of a target user.

\noindent \textbf{Metric-based MI.}~~ \citet{shejwalkar2021membership} also propose a metric-based MI that uses a threshold value of prediction loss to decide the membership status of a single data sample. Then, based on the membership status of each sample, a majority vote over the samples is used to infer the membership status of the data owner. Compared to learning-based MI, metric-based MI does not need large computational resources to train multiple shadow models, but it still requires the proxy datasets to fine-tune the value of the threshold.

\subsection{\method} 
Traditional MI methods, e.g., learning-based and metric-based MI, are developed from the perspective of an attacker, i.e., an attacker is given as much information as possible to launch MI attacks to breach the membership privacy of a user. Thus, it is impractical to leverage the existing MI methods to protect users' data because the information required for them is rarely available to a user. To solve this obstacle, we propose \method, i.e., a backdoor-based membership inference approach, which requires only black-box access to any NLP models, such as LLMs and others. Our \method leverages a key observation that a backdoored model will behave very differently from a clean model when predicting inputs with backdoor triggers. Thus, a user can previously add backdoor triggers to protect her data so that at a later stage, she can infer whether her data was used to train the NLP model without authority by identifying whether the target model is backdoored. \method has two important steps:

\noindent \textbf{Generate backdoored text.}~~ The data owner follows the standard and the most widely used backdoor attacks proposed in \cite{gu2019badnets} to generate backdoored text. Specifically, an original sample $(\bm{x},y)$ of the data owner has the original feature $\bm{x}$ and its true label $y$. The data owner inserts the backdoor trigger into $\bm{x}$, which will be modified to $\bm{x}_t$. Then, a target label $y_{t}$ ($y_{t} \neq y$) is assigned to $\bm{x}_t$, which is a different label than the ground-true label $y$. Finally, the data owner generates a backdoored text sample $(\bm{x}_t,y_t)$. To backdoor the target model, the data owner may need to generate a few backdoored samples. In the experiments, we will investigate how many backdoored samples are required for a single user. In this paper, we systematically investigate three types of backdoor triggers, i.e., character-level, word-level, and sentence-level triggers, that the data owner can leverage to protect her private character, word, and sentence, respectively. For example, a user can leverage different triggers to protect her private text in Table~\ref{table:trigger_usage}. 

\noindent \textbf{Verification Framework.}~~ The data owner infers whether her sensitive text information was used to train the NLP model by verifying whether the NLP model is backdoored. Inspired by \cite{hu2022membership}, we leverage hypothesis testing to implement the verification. Specifically, let $f(\cdot)$ be the text classification model. We define a null hypothesis $\mathcal{H}_{0}$ and an alternative hypothesis $\mathcal{H}_{1}$ as follows:
\begin{equation}
    \begin{array}{l}
{\mathcal{H}_0}:  \Pr \left(f(\bm{x}_{t}) = {y_t} \right) \le \beta, \\
{\mathcal{H}_1}: \Pr \left(f(\bm{x}_{t}) = {y_t} \right)  > \beta,
\end{array}
\end{equation}
where $\Pr \left(f(\bm{x}_{t}) = {y_t} \right)$ represents the backdoor attack success probability of the target NLP model, and $\beta$ represents the backdoor attack success probability of a clean model. The null hypothesis $\mathcal{H}_{0}$ represents that there are no significant differences between the behavior of the target model and a clean model. On the contrary, $\mathcal{H}_{1}$ represents that the target NLP model behaves significantly differently from a clean model. If the data owner can reject the null hypothesis $\mathcal{H}_{0}$ with statistical guarantees, she can claim that her text was used to train the target NLP model. We use the attack success rate (ASR) of the target model to conduct the hypothesis test. Specifically, if the ASR of the model is higher than a threshold, the data owner would consider her data was used to train the model. 
% How to calculate the threshold can refer to Appendix~\ref{Appendix:test}.
% To counter the class imbalance problems that are practical among real-world datasets, different from selecting the random guess probability as $\beta$ in \cite{hu2022membership}, we propose the following theorem with class-wise $\beta$ for a higher MI successful rate. 
We leverage the Theorem \ref{theorem:t-test} form \cite{hu2022membership} to obtain the thresholding ASR.

% \begin{restatable}{theorem}{ttest}\label{theorem:t-test}
% Given a target model $f(\cdot)$ and the number of classes $C$ in the classification task, with the number of queries to $f(\cdot)$ at $m$ ($m \geq 30$), if the backdoor attack success rate (ASR) $\alpha$ of $f(\cdot)$ satisfies the following formula:
% \begin{gather*}
%     \sqrt {m - 1}  \cdot (\alpha  - \beta ) - \sqrt {\alpha - {\alpha^2}}  \cdot {t_\tau } > 0,
% \end{gather*}
% the data owner can reject the null hypothesis $\mathcal{H}_{0}$ at the significance level $1-\tau$, where $\beta=\frac{1}{C}$ and ${t_\tau }$ is the $\tau$ quantile of the t distribution with $m-1$ degrees of freedom.
% \end{restatable}
\begin{restatable}[Finding ASR threshold via T-test]{thm}{theorem:t-test}
\label{theorem:t-test}
Given a target model $f(\cdot)$ and the number of classes $C$ in the classification task, with the number of queries to $f(\cdot)$ at $N$ ($N \geq 30$), if the backdoor attack success rate (ASR) $\alpha$ of $f(\cdot)$ satisfies the following formula:
\begin{gather*}
    \sqrt {N - 1}  \cdot (\alpha  - \beta ) - \sqrt {\alpha - {\alpha^2}}  \cdot {t_\tau } > 0,
\end{gather*}
the data owner can reject the null hypothesis $\mathcal{H}_{0}$ at the significance level $1-\tau$, where $\beta$ is a certain threshold and ${t_\tau }$ is the $\tau$ quantile of the t distribution with $N-1$ degrees of freedom.
\end{restatable}
% $\beta=\frac{1}{C}$ and 

\noindent \textbf{Verification Threshold $\beta$.} There are two common training settings: i) training from scratch with a random-initialized backbone and ii) fine-tuning from an open-source pre-trained backbone. In this paper, different from \cite{hu2022membership}, we focus on the fine-tuning case, which provides more prior for conducting efficient MI. For the first case, since we have no prior knowledge of the random-initialized backbones, setting the threshold $\beta$ to be the random-guess level $\beta = \frac{1}{C}$ can be an astute and model-agnostics policy that allows transferability. Since for trained clean models, $\Pr \left(f(\bm{x}_{t}) = {y_t} \right) < \frac{1}{C}$ holds anyway. However, such thresholds might not be efficient for the second case when we have additional information and access to the pre-trained model $\theta_0$. Support we denote the fine-tuned model $\theta_{u}$ on user data $D_u$, we set the $\beta=\text{Pr}[f_{\theta_0}(x_t)=y_t]$ by observing that the following inequation holds: 
\begin{equation}
    \beta(\theta_0)= \text{Pr}[f_{\theta_0}(x_t)=y_t]<\frac{1}{C}<\text{Pr}[f_{\theta_u}(x_t)=y_t],
\end{equation}
where the first inequality sign suggests that the pre-trained model turns to predict the backdoored sample $x_t$ to its original label $y(x)$, and the second one suggests that fine-tuning the backdoored data will lead to ASR improvement to a level that is better than random guessing. Since for the first training setting, the randomly initialized nets will yield random guessing prediction, we thus can unify both two training settings into a single framework. We prove this threshold is more efficient in Theorem \ref{theorem:ours}. 

\begin{restatable}[Verifying and Marking Efficiency of $\beta(\theta_0)$]{thm}{ourtheorem}
\label{theorem:ours}
    Given the pre-trained weight $\theta_0$, setting a $\theta_0$-conditioned $\beta$, i.e., $\beta= \text{Pr}[f_{\theta_0}(x_t)=y_t] $, can lead to more efficient verifying and also less marking ratio, compared to setting the random-guessing threshold $\beta_{\text{rand}}=\frac{1}{C}$. 
\end{restatable}

% \ourtheorem*
\proof 
We first prove the verifying efficiency under the setting that a model trained on marked data sets with a fixed marking rate. We simplify the notation of attack successful rate of the trained model as $\alpha = \Pr \left(f_{\theta_u}(\bm{x}_{t}) = {y_t} \right)$. Given the original benign accuracy of pre-trained model $\theta_0$ as $\operatorname{Pr}\left[f_{\theta_0}\left(x_t\right)=y_t\right] $, we have the possibility of an event that ASR is larger than $\beta$ as 
% {\mathcal{H}_0}:  \Pr \left(f(\bm{x}_{t}) = {y_t} \right) \le \beta, \\
% {\mathcal{H}_1}: \Pr \left(f(\bm{x}_{t}) = {y_t} \right)  > \beta,
\begin{equation}
\begin{aligned}
  \operatorname{Pr}[\alpha \geq \beta ] &= \operatorname{Pr}[\alpha \geq \operatorname{Pr}\left[f_{\theta_0}\left(x_t\right)=y_t\right] ]  \\
  &= \operatorname{Pr}[\alpha \geq \frac{1}{C} \text{ or} \frac{1}{C} > \alpha > \beta ] \\ 
  &= \operatorname{Pr}[\alpha \geq \frac{1}{C}]+  \operatorname{Pr}[\frac{1}{C} > \alpha \geq  \beta ] \\
  & \geq \operatorname{Pr}[\alpha \geq \frac{1}{C}] \\ 
  & =\operatorname{Pr}[\alpha \geq \beta_{\text{rand}}]
\end{aligned}
\end{equation}
Recall that we set the null hypothesis ${\mathcal{H}_0}:  \alpha \le \beta$ and the alternative hypothesis ${\mathcal{H}_1}: \alpha  > \beta$, we thus can conclude that for the alternative hypothesis, we now have the following equation for rejecting the null hypothesis and accepting the alternative one: 
\begin{equation}
\small 
    \operatorname{Pr}[\text{Accepting }\mathcal{H}_1 \text{ with }\beta]  \geq \operatorname{Pr}[\text{Accepting }\mathcal{H}_1 \text{ with }\beta_{\text{rand}}] 
\end{equation}
We thus prove the verifying efficiency since we can more easily verify membership with the lower false negative ratio for a fixed marking rate. For marking efficiency, increasing the number of marked samples can lead to an increase in ASR $\alpha$. Since we now have a lower $\beta$ threshold, we thus require fewer marking samples to achieve the same membership inference performance as the original one. We conclude that our $\beta$ setting policy can achieve better verifying efficiency with lower marking rates. 
\endproof

\header{Remark} We set $N=30$, $\tau=0.05$ following \cite{hu2022membership}. To obtain $\text{Pr}[f_{\theta_0}(x_t)=y_t]$, we take a zero-shot classifier that is built with a pre-trained model $f_{\theta_0}$ as an embedding extractor for both input text and label text, and use the cosine similarity metric for making the prediction. 

\begin{table*}[thbp]
  \centering

  \resizebox{0.8\linewidth}{!}{
    \begin{tabular}{c l c c c c}
      \toprule
      \textbf{Dataset} & \textbf{Methods} & \textbf{Roberta} & \textbf{Albert-v2} & \textbf{GPT2} & \textbf{T5} \\
      \midrule
      \multirow{3}{*}{
      GenderBias
      } & T-SMIA & 33\% $\pm$ 0\% & 33\% $\pm$ 1\% & 33\% $\pm$ 9\% & 33\% $\pm$ 6\% \\
       & L-UMIA & 33\% $\pm$ 5\% & 33\% $\pm$ 3\% & 43\% $\pm$ 1\% & 38\% $\pm$ 2\% \\
       & \textbf{TextMarker} & \textbf{100\% $\pm$ 7\%} & \textbf{79\% $\pm$ 0\%} & \textbf{100\% $\pm$ 1\%} & \textbf{79\% $\pm$ 0\%} \\
      \midrule
       \multirow{3}{*}{
      IMDB
      } & T-SMIA & 58\% $\pm$ 7\% & 42\% $\pm$ 16\% & 48\% $\pm$ 0\% & - \\
       & L-UMIA & 56\% $\pm$ 0\% & 33\% $\pm$ 0\% & 67\% $\pm$ 1\% & 33\% $\pm$ 0\% \\
       & \textbf{TextMarker} & \textbf{95\% $\pm$ 16\%} & \textbf{78\% $\pm$ 0\%} & \textbf{79\% $\pm$ 5\%} & \textbf{33\% $\pm$ 0\%} \\
      \midrule
        \multirow{3}{*}{
      SST-5
      } & T-SMIA & 33\% $\pm$ 0\% & 33\% $\pm$ 0\% & 34\% $\pm$ 0\% & 33\% $\pm$ 40\% \\
       & L-UMIA & 57\% $\pm$ 0\% & 33\% $\pm$ 0\% & 52\% $\pm$ 0\% & 33\% $\pm$ 0\% \\
       & \textbf{TextMarker} & \textbf{90\% $\pm$ 1\%} & \textbf{80\% $\pm$ 8\%} & \textbf{100\% $\pm$ 6\%} & \textbf{52\% $\pm$ 0\%} \\
      \midrule
       \multirow{3}{*}{
     Trec
      } & T-SMIA & 33\% $\pm$ 1\% & 33\% $\pm$ 0\% & 33\% $\pm$ 2\% & 33\% $\pm$ 1\% \\
       & L-UMIA & 34\% $\pm$ 14\% & 38\% $\pm$ 11\% & 37\% $\pm$ 14\% & 33\% $\pm$ 0\% \\
       & \textbf{TextMarker} & \textbf{90\% $\pm$ 14\%} & \textbf{90\% $\pm$ 0\%} & \textbf{100\% $\pm$ 0\%} & \textbf{79\% $\pm$ 0\%} \\
      \bottomrule
      \end{tabular}
  }
  \caption{Performance comparison in terms of F1-score ($\uparrow$) for MI between our method and two MI baselines on various datasets and model architectures. For the baselines T-SMIA and L-UMIA, the proportion of true training and testing set used for proxy data construction is 50\%. For TextMarker, the total marking ratio $R$ is $\sim$ 7\%.}

  \label{tab:more_results}
\end{table*}

\begin{figure*}[!t]
    \centering
    \includegraphics[width=0.9\linewidth]{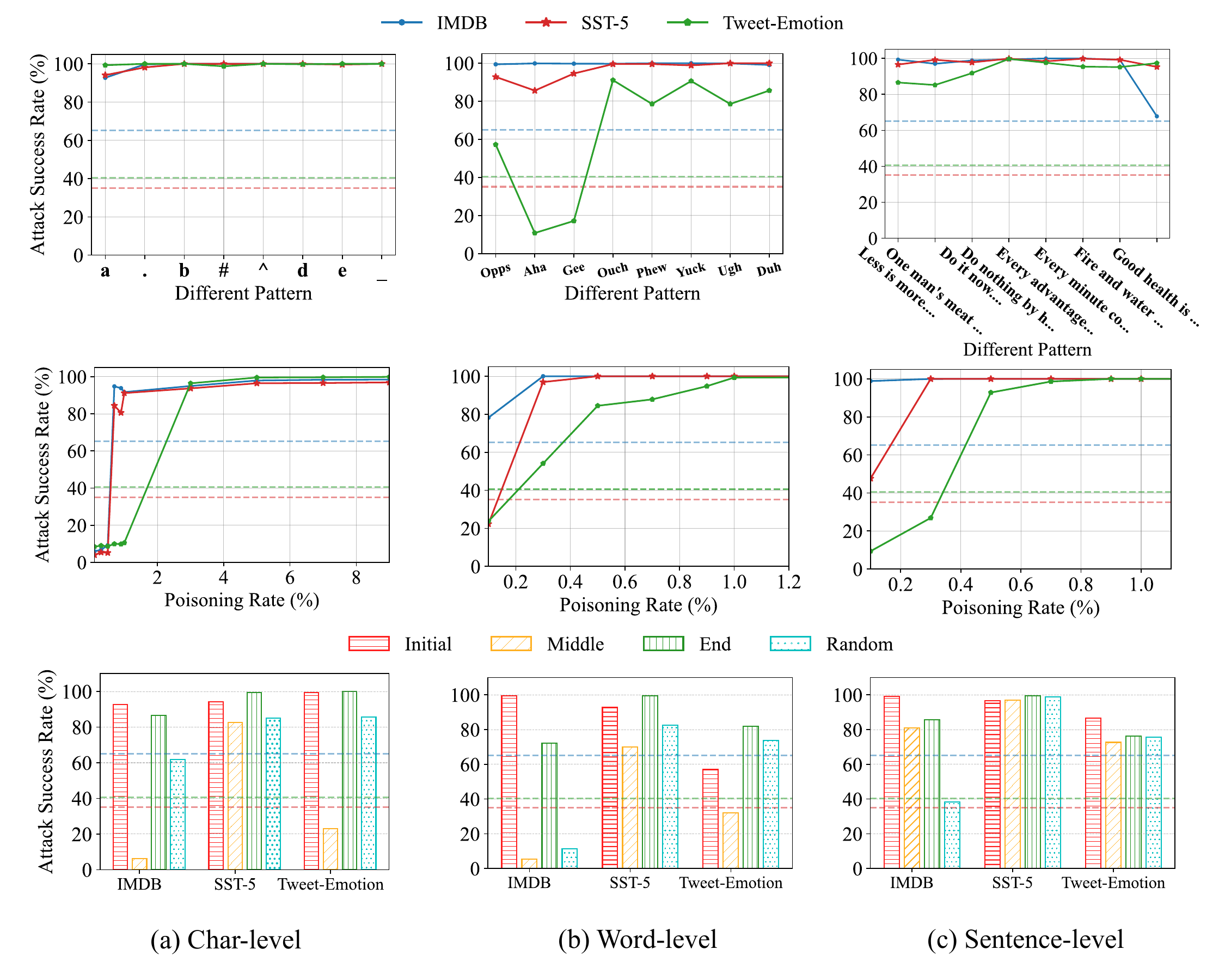}
    \vspace{-5pt}
    \caption{
    The sensitivity study of trigger configurations. The dotted lines indicate the ASR threshold for MI. The ASR above the threshold indicates a successful MI. 
    % Understanding the impact of trigger patterns/poison rate/trigger location  in \method. In each plot, the $x$ axis represents the different parameters, and the $y$ axis represents ASR. 
% % The dotted line indicates the ASR threshold that allows successful membership inference.
% The dotted lines indicate the ASR threshold that allows successful membership inference (a blue line for IMDB, a red line for SST-5, and a green line for Tweet-Emotion).
}
    \label{fig:sensitivity}
\end{figure*}

\section{Experiments}
\label{sec:exp}

% In this section, we first introduce the experiment settings. Then, we show the effectiveness and efficiency of our method and compare it with the existing MI approaches. Lastly, we first conduct robustness analysis to investigate how \method performs under a clean-fine-tuning-based watermark removal technique. Then, we conduct a robustness and sensitivity analysis to show how different factors can affect our method. 

\subsection{Experimental Settings}
% \noindent \textbf{Datasets.} To evaluate our methods, we conduct experiments using three text sentiment analysis datasets, including IMDB (2 classes)\cite{imdb}, Twitter-Emotion (4 classes)\cite{emotion}, and SST-5 (5 classes)\cite{sst5}. For detailed statistics of all datasets, please refer to Table \ref{tab:dataset} in Appendix \ref{Appendix1: datastatistics}.
% \paragraph{Datasets and Target model.}~~Statistics of the datasets are presented in Table \ref{tab:dataset}. We mainly focus on the sentiment analysis task in this paper. The datasets can be downloaded from the Hugging Face website\footnote{\href{https://huggingface.co/datasets}{https://huggingface.co/datasets}}. 
% % Pre-training on large-scale data and then performing fine-tuning on downstream tasks have become a dominant method in the domain of NLP, which achieves state-of-the-art performance. 
% % As for the target model, 
% % we mainly focus on the commonly used Transformer-based architecture. 
% We consider the \textit{bert-base-uncased} as the target model, which has 24 transformer encoders. 
% , which has 24 transformer layers, with 768 hidden units and 12 self-attention heads.
\paragraph{Datasets and target models.}
% Statistics of the datasets are presented in Table \ref{tab:dataset}. In this paper, we conduct experiments on both the sentiment classification task () and natural language inference task (). 
 We conduct experiments on a variety of tasks, including sentiment classification (IMDB, SST-5, Tweet-Emotion), natural language inference (NLI) (MultiNLI-Fiction, MultiNLI-Government), gender bias classification (MDGenderBias-wizard subset), and question classification (TREC)\footnote{\href{https://huggingface.co/datasets}{https://huggingface.co/datasets}}. We include various LLM architectures for extracting text embedding, including encoder-only LMs (\textit{BERT, Roberta, DistilBERT, Albert-v2}), causal-decoder LMs (\textit{DistilGPT2}), and encoder-decoder LMs (\textit{T5}). An additional linear layer is used to build the classification head. The \texttt{bert-base-uncased} \cite{devlin2018bert} is selected by default with 24 transformer encoders. 
  % The full statistics of the datasets are presented in the App. \ref{Appendix: training}. 

% \paragraph{Datasets and Target Models.}~~We conduct evaluation on various datasets and model architectures. Statistics of the datasets are presented in Table \ref{tab:dataset}. The datasets encompass various text classification tasks and can be downloaded from the Hugging Face website\footnote{\href{https://huggingface.co/datasets}{https://huggingface.co/datasets}}. We consider several language models as target models, including \textit{bert-base-uncased}, \textit{Roberta}, \textit{DistilBERT}, \textit{Albert-v2}, \textit{DistilGPT2}, and \textit{T5}, each having a different number of transformer encoders and varying architectures.

% \noindent \textbf{ASR threshold for MI.}~~ To reject the null hypothesis $\mathcal{H}_{0}$, i.e., claiming the target model is backdoored, the data owner does not need the ASR of the target model to be perfect ({e.g., 100\%} ) but requires the ASR to be larger than a threshold. We set the significance level of the hypothesis test at 0.05, which is the common practice in statistical hypothesis testing~\cite {craparo2007significance}. 
% Based on Theorem~\ref{theorem:t-test} in Appendix \ref{Appendix:test}, the threshold of ASR to reject $\mathcal{H}_{0}$ is calculated as 65.1\%, 40.5\%, and 35.1\% for IMDB, SST-5, and Tweet Emotions, respectively. 

\noindent \textbf{Baselines and metrics.}~~ We compare our proposed \method against the two existing MI methods, i.e, T-SMIA \cite{nasr2019comprehensive} (metric-based MI) and L-UMIA \cite{shejwalkar2021membership} (learning-based MI). As the MI problem is a binary classification problem, we use accuracy, recall, precision, and F1 score following \cite{hu2021membership}.

\header{Data owner and trigger} We study the scenario of both single data owners and multiple data owners. For the latter case, data owners are forced to have different trigger patterns. For the different levels of backdoor triggers, we construct the corresponding trigger dictionary. Following \cite{chen2021badnl, sun2020natural}, we use neutral triggers to avoid negatively impacting the original model's utility. Specifically, we select triggers of English letters and punctuation marks for the character-level method, {e.g.,} `a' and `\#'. For the word-level method, we use the triggers of modal particle, {e.g.,} `Ops'. For the sentence-level method, we have select triggers of English idiom, {e.g.,} `Less is more'. After choosing one trigger from a dictionary, we randomly determine the size and location of the trigger location and then insert it or replace the original text. 

\begin{table*}[thbp]
\centering
\resizebox{.8\linewidth}{!}{
\begin{tabular}{ccccccccc} 
\toprule
\multirow{2}{*}{\textbf{Datasets}} & \multicolumn{4}{c}{$\beta=\frac{1}{C}$}                                                                                & \multicolumn{4}{c}{$\beta=\text{Pr}[f_{\theta_0}(x_t)=y_t]$}                                    \\ 
\cmidrule{2-9}
                                   & \textbf{Clean Utility} & $\min(R) \downarrow$ & $\alpha*$ & \textbf{ASR} & \textbf{Clean Utility} & $\min(R) \downarrow$ & $\alpha*$ & \textbf{\textbf{ASR}}  \\ 
\midrule
Fiction                            & 80.79               & 0.10\%                                                                & 49.11     & 96.47        &         81.15             &                         0.01\%       &    9.59           &           42.22                  \\
Government                         & 83.80              & 0.15\%                                                                & 49.11     & 97.89        &       83.81                    &                        0.02\%     &    39.42      &     58.31              \\
\bottomrule
\end{tabular}
}
\caption{The comparison between \method with two different $\beta$ policies of applying our Sentence-level trigger on the natural language inference task (Multi-NLI dataset). Our method is more efficient in conducting MI with much less minimum marking ratio $\min(R)$. 
% Our method can achieve MI (Membership Inference) successfully with a minimum marking ratio of ~0.1\% while maintaining the benign performance of the market model. The fiction and government subset of Multi-NLI are selected for demonstration. The minimum masking ratio is determined with a grid search.
}
\label{tab:inference_task_results}
\end{table*}

\subsection{Effectiveness and Efficiency}
% \header{Effectiveness}
We compare our proposed method with the traditional MI methods. For each dataset, we select 50 member users from the training dataset and 50 non-member users from the testing dataset of the target model. Each user randomly selects a trigger among the three types of triggers and the configuration. Note that each owner has a unique pattern and trigger label to avoid causing conflicts in their own membership inference purposes. 
% In this experiment, we let different users choose different trigger patterns, target labels, and poison rates for their own membership inference purposes.
% Each user has 5 data samples.
% For our proposed backdoor-based approach, each user chooses different trigger patterns, and target labels to craft a different number of text samples. 
% The trigger generation settings are described in App. \ref{} with $R$ (\%) as the marking rate for each data owner. 
For the learning-based MI and the metric-based MI methods, we assume they can directly access half of the rest of the training samples and testing samples of the target model. Note that in the previous work \cite{shejwalkar2021membership}, two traditional MI methods only have access to the proxy datasets, which are disjoint from the training dataset and the testing dataset of the target model. In our setting, the learning-based MI and the metric-based MI methods can achieve stronger performance than the original ones.
% In Appendix \ref{Appendix:more_compare}, we have further evaluated and compared our method against the learning-based MI and the metric-based MI methods when they can access different portion of the training and testing dataset of the target mode.
Table~\ref{tab:more_results} shows that the proposed method is superior to the two existing MI methods, even though they can access more raw training data and testing data of the target model. One of the explanations behind the success of our proposed method is that we design a ``one for one'' method, i.e., each user generates their own backdoored text, which injects a unique backdoor into the target model. Then, each user can use a unique trigger to verify the membership status of their data. In contrast, the learning-based and metric-based MI methods are ``one for all'' methods, i.e., they try to learn a single classifier or use a single threshold to decide the membership status of all users. Thus, they may not generalize well and can not conduct MI precisely. 
% Moreover, we test the benign accuracy of models trained on marked datasets and find that our method will not affect the utility. 
For efficiency, we further demonstrate that \method is more sample-efficient than the previous approach \cite{hu2022membership} that takes the random guessing threshold to compute the boundary ASR. For comparison, we show the minimum marking ratios and also the boundary ASR for conducting successful MI. Table \ref{tab:inference_task_results} shows that our threshold policy can be $\sim 10 \times $ more efficient in marking ratio.

\subsection{Robustness and Sensitivity}
% \header{Robustness analysis}
% Under an adverse setting, the model trainer might conduct additional fine-tuning with clean data similar to the user's data in distribution, which is a common watermark removal technique \cite{pang2023backdoor}. In this setting, we seek to explore how resilient our method is under different strengths of clean fine-tuning. Specifically, we conduct experiments with additional fine-tuning epochs for the marked model on a leave-out clean training subset. The results in Table \ref{tab:robustness} indicate that the injected watermark remains notably robust even with additional fine-tuning epochs. Regardless of the number of additional clean samples introduced, the MI F1 score remains high, showcasing the resilience of our method. However, a slight decrease in the F1 score is observed with the increase of a clean fine-tuning set, which might be caused by a phenomenon known as ``Catastrophic Forgetting'' \cite{french1999catastrophic} in deep neural networks. Despite this, our method demonstrates promising robustness against the fine-tuning-based watermark removal method, indicating its potential for effective utilization even in adversarial settings with watermark removal. 

\header{Robustness analysis}
We explore the resilience of our method under different strengths of clean fine-tuning, a common watermark removal technique \cite{pang2023backdoor}. Experiments with additional fine-tuning epochs on a leave-out clean training subset show that the injected watermark remains notably robust (Table \ref{tab:robustness}). The MI F1 score remains high, showcasing the method's resilience despite a slight decrease due to "Catastrophic Forgetting." Our method demonstrates promising robustness against fine-tuning-based watermark removal, indicating its potential even in adversarial settings.

\begin{table}[thbp]
  \centering

\resizebox{\linewidth}{!}{
  \begin{tabular}{ccccc}
    \toprule
    \textbf{$+$ Clean Fine-tuning} & \textbf{Accuracy ($\uparrow$)} & \textbf{Recall ($\uparrow$)} & \textbf{Precision ($\uparrow$)} & \textbf{F1-score ($\uparrow$)} \\
    \midrule
   $\times $ & 90.0\% & 90.0\% & 91.7\% & 89.9\% \\
    
   + 30\% $\mathcal{D}_{\text{clean}}$ & 90.0\% & 90.0\% & 91.7\% & 89.9\% \\
    
    + 40\% $\mathcal{D}_{\text{clean}}$ & 83.3\% & 83.3\% & 87.7\% & 82.7\% \\
    
    + 50\% $\mathcal{D}_{\text{clean}}$& 76.7\% & 76.7\% & 78.8\% & 76.3\% \\
    \bottomrule
    \end{tabular}
}
\caption{The MI Performance of \method against the fine-tuning-based watermark removal method.}
  \label{tab:robustness}
\end{table}

% \header{Sensitivity analysis} We conduct detailed sensitivity studies to investigate how different factors affect the performance of \method. Specifically, we evaluate the effect of trigger patterns, trigger locations, and poison rates.
% As for the pattern and trigger location, we observe that most of the patterns considered are effective for successful MI. However, the results also show that some patterns like ``Aha'' are less effective than others, indicating the importance of choosing the right patterns. As for the location, we evaluate {four trigger locations}, \textit{initial, middle, end, and random} (\ie, randomly determine where to insert), for all three types of triggers. The results in Figure \ref{fig:sensitivity} show that applying the random location strategy in general leads to better performance than others, which can be explained as a way with in-explicit data argumentation that helps the model learn the backdoor correlation. On the marking rates, the results in Figure \ref{fig:sensitivity} show that, for all three types of triggers, when the poisoning rate is high enough ($>$2\%), the data owner can have the ASR of the target model higher than the threshold. 
\header{Sensitivity analysis}
We conduct sensitivity studies to investigate the impact of trigger patterns, locations, and poison rates on the performance of \method. Figure \ref{fig:sensitivity} shows that most patterns are effective for successful MI, with some exceptions like ``Aha''. Applying the random location strategy generally leads to better performance than fixed locations (initial, middle, end). The results also demonstrate that when the poisoning rate exceeds 2\%, the data owner can achieve an ASR higher than the threshold for all three trigger types.

\section{Conclusion}
\label{sec:conclusion}
In this paper, we propose \method, a text data watermarking approach leveraging backdoor-based membership inference for users to identify the unauthorized usage of their private text. Our proposed \method is superior to previous methods, with an efficient threshold technique, requiring far fewer marking samples to achieve membership inference. Through extensive experiments, we evaluate the effectiveness of \method across various datasets and model architectures. 
% Future works can further adapt \method to in-context learning. 
Despite the effectiveness, one of the limitations is that currently we only focus on the text classification training setting. An important future direction is to adapt \method to in-context learning setting. 

\clearpage

\bibliography{reference}

% \appendix
% \input{sections/appendix/app}

\end{document}